\documentclass[12pt]{article}
\usepackage{psfig}
\usepackage{amsmath}
\usepackage{amssymb}

\usepackage[dvips]{color}

\newcommand{\Black}{\color [rgb]{0,0,0}}

\newcommand{\Brown}{\color [rgb]{0.4,0.1,0.1}}

\def\TL{\hfil$\displaystyle{##}$}
\def\TR{$\displaystyle{{}##}$\hfil}

\def\TT{\hbox{##}}
\def\seqalign#1#2{\vcenter{\openup1\jot
  \halign{\strut #1\cr #2 \cr}}}

\def\fixit#1{}

\def\mop#1{\mathop{\rm #1}\nolimits}

\def\vol{\mop{vol}}

\def\overleftrightarrow#1{\vbox{\ialign{##\crcr
     $\leftrightarrow$\crcr\noalign{\kern-0pt\nointerlineskip}
     $\hfil\displaystyle{#1}\hfil$\crcr}}}

\def\lsim{\mathrel{\mathstrut\smash{\ooalign{\raise2.5pt\hbox{$<$}\cr\lower2.5pt\hbox{$\sim$}}}}}
\def\gsim{\mathrel{\mathstrut\smash{\ooalign{\raise2.5pt\hbox{$>$}\cr\lower2.5pt\hbox{$\sim$}}}}}
\def\sqr#1#2{{\vcenter{\vbox{\hrule height.#2pt
         \hbox{\vrule width.#2pt height#1pt \kern#1pt
            \vrule width.#2pt}
         \hrule height.#2pt}}}}
\def\square{\mathop{\mathchoice\sqr86\sqr76\sqr{3.75}4\sqr56\,}\nolimits}
\def\href#1#2{#2}

\def\lbldef#1#2{\expandafter\gdef\csname #1\endcsname {#2}}
\def\eqn#1#2{\lbldef{#1}{(\ref{#1})}%
\begin{equation} #2 \label{#1} \end{equation}}
\def\eqalign#1{\vcenter{\openup1\jot
    \halign{\strut\span\TL & \span\TR\cr #1 \cr
   }}}

\begin{document}
\baselineskip=15.5pt \pagestyle{plain} \setcounter{page}{1}
%\renewcommand{\thefootnote}{\fnsymbol{footnote}}
%--------+---------+---------+---------+---------+---------+---------+
%Title page

\begin{titlepage}

\begin{flushright}
UCB-PTH-05/21\\
LBNL-58216\\
hep-th/0508238
\end{flushright}
\vfil

\begin{center}
{\huge Investigating the Stability of } \vskip0.5cm {\huge
 a Nonsupersymmetric Landscape}
\end{center}

\vfil
\begin{center}
{\large Yoon Pyo Hong$^{1}$ and Indrajit Mitra$^{1,2}$}
\end{center}

$$\seqalign{\span\TL & \span\TT}{
{}^1 & Berkeley Center for Theoretical Physics and Department of
Physics \cr\noalign{\vskip-1.5\jot} & University of California,
Berkeley, CA 94720-7300 \cr {}^2 & Theoretical Physics Group,
Lawrence Berkeley National Laboratory \cr\noalign{\vskip-1.5\jot}
& Berkeley, CA 94720-8162 }$$ \vfil

\begin{center}
{\large Abstract}
\end{center}

\indent We investigate the classical stability of
non-supersymmetric Freund--Rubin compactifications of Type IIB
string theory on a product of three-dimensional Einstein spaces
$A_3 \times B_3$ with both NS-NS and R-R three-form fluxes turned
on through $A_3$ and $B_3$, and a zero axion. This results in a
three parameter family of $AdS_4$ vacua, with localized sources
such as anti-three-branes or orientifold planes required to cancel
the R-R four-form tadpole. We scan the entire space of such
solutions for perturbative stability and find that
anti-three-branes are unstable to a Jeans-like instability. For
orientifold compactifications, we derive a precise criterion which
the three-dimensional Einstein spaces have to satisfy in order to
be stable.

\vfil
\begin{flushleft}
August 2005
\end{flushleft}
\end{titlepage}
\newpage
%--------+---------+---------+---------+---------+---------+---------+
%Body
\section{Introduction}
\label{Introduction}

String theory compactified down to four dimensions has a richness of
vacua. Studies of these vacua in the context of different string
theories are by now a well-developed industry (see e.g. \cite{KKLT,
  DGKT, DGKTTwo, Oliver, FFix, Conlon, Vijay, Acharya, Curio, BeckOne,
  BeckTwo, BA, STW, AMOne, AMTwo, Krause}). Counting the number of these vacua has been the focus of many
investigations --- the considerations of Bousso and
Polchinski \cite{BoussoPol}, followed up by the more
sophisticated and detailed investigations for vacua preserving ${\cal
  N}=1$ supersymmetry by Douglas and collaborators
\cite{Douglas, DA, DD}. This plethora of vacua leads to an
embarrassment of richness in making contact with four-dimensional
physics. It is therefore interesting to ask how many vacua in this
landscape \cite{Land} remain perturbatively stable once
supersymmetry is broken. In this paper, we examine a class of
compactifications of Type IIB string theory which have
non-supersymmetric four-dimensional AdS vacua, and perform a
stability analysis. (For somewhat related work, see
\cite{DNONSUSY}. For a study of the stability of compactifications
with de Sitter vacua, see e.g. \cite{BDM,KPZ}.)

Ideally, one would like to start with a model with a large number
of supersymmetric vacua, break supersymmetry, and then perform a
counting of the number of stable vacua. This is not what we do.
Instead, we consider a set-up which is non-supersymmetric from the
start. What we have in mind is a Freund--Rubin compactification
\cite{FR} of Type IIB on a six-dimensional space $X_6$ which is a
direct product of three-dimensional positively curved Einstein
spaces $A_3 \times B_3$, with 3-form fluxes threading both $A_3$
and $B_3$. In these solutions the radii of curvature of $A_3$ and
$B_3$ increase with 3-form fluxes. We consider large fluxes where
classical supergravity can be trusted. In the absence of
supersymmetry, the small radius limit will receive large
corrections.\footnote{Even in
  the large radius limit one could
still worry that close to the orientifold planes (which we
introduce shortly) there are significant corrections to the
geometry. Moreover, our very geometric approach to objects that
are intrinsically non-geometric might not seem rigorous. We shall
gloss over both of these subtleties by hoping that at low
energies, our analysis would be nearly correct.} To our knowledge,
this class of solutions appeared first in \cite{EvaModuli}. In
this paper we examine the perturbative stability of these
solutions.

 There are at least three motivations to be interested in the study of
 the stability of such compactifications:
 \begin{itemize}
 \item These are the simplest compactifications of IIB string theory
 with non-trivial three-cycles through which one can thread $H_3$ and
 $F_3$ flux. It is therefore very interesting to ask how many of the
 non-supersymmetric vacua are classically stable.

 \item The flux compactifications under study here arise as
 near-horizon limits of stacks of NS5 and D5 branes wrapping $A_3$ and $B_3$. They are therefore similar to the KKLT and KKLMMT models
 \cite{KKLT, KKLLMMT} which also have 5-branes wrapping three-cycles in a Calabi-Yau. As such, the set-up
 considered here could be viewed as a toy-model of the more
 sophisticated Calabi-Yau constructions.{\footnote {Recall that in the
 de Sitter compactifications mentioned above, the first step is to
 obtain an AdS vacuum. One then adds anti-three-branes which more
 than cancel the negative cosmological constant to obtain a meta-stable
 de Sitter vacuum.}} This toy-model might provide some mileage in
 constructing putative three-dimensional field theory duals,
 especially since the brane configurations are known here.

 \item  Finally, since the issue of stability of product space
 compactifications with
 fluxes threading the individual parts of the product has (to our
 knowledge) not been carried out (see \cite{Campbell} however for a
 closely related set-up), the exercise is interesting by itself in
 getting an idea of when such compactifications are
 unstable.

 \end{itemize}

 When one desires a 4-d AdS vacuum in some theory such as IIB, one
 still has to choose a particular 6-d compact space, $X_6$. In order
 to be able to turn on 3-form fluxes, one requires that $X_6$ have some number
 of three-cycles. The simplest choice is then a direct product of three-manifolds.
 Once one has made a choice of $X_6$, each
 of the four dimensional anti-de Sitter vacua are differentiated by
 the amounts of 3-form flux turned on through each three-cycle.
 In general these would be four integers $(n_1,n_2,q_1,q_2)$,
 with $n_1$, $n_2$ specifying the amount
 of NS-NS flux through $A_3$ and $B_3$, respectively, and $q_1$, $q_2$ the R-R
 flux. However, for simplicity, we shall choose an ansatz with
 no axion turned on. This yields a constraint relating the four
 integers. There is also a $C_4$ tadpole constraint, which is Gauss's law for 3-brane (or
 $H_3 \wedge F_3$) charge in the compact space. A non-zero $H_3 \wedge
 F_3$ necessitates the introduction of localized objects which carry
 3-brane charge. The various options in IIB string theory are
 D3-branes and O3-planes (or $\overline {\mathrm{D3}}$ and $\overline{\mathrm{O3}}$,
 depending on the relative orientation of $H_3$ and
 $F_3$), which are point-like in $X_6$, or $p$--dimensional branes wrapped
 on $(p-3)$-cycles in $X_6$. For simplicity we consider only the first two
 options. We find that if we denote the number of localized objects needed to
 cancel the $C_4$ tadpole by $t$, we have a three parameter family of
 $AdS_4$ vacua which are conveniently labelled by $t$, $\alpha = {n_1/n_2}$, and $g_s = e^{\phi}$,
 whose background value we shall see is
 ${{\alpha n_2}/q_2}$. The entropy of these solutions (which is
 related to the number of AdS vacua) depends on fluxes, and the relationship can be computed from the value of the four-dimensional
 cosmological constant using the standard argument of Witten and
 Susskind \cite{SW}. The result is that the entropy is proportional to $t^2$. (Recall that for
 supersymmetric vacua constrained to obey the $C_4$ tadpole condition,
 the entropy scales as $t^{b_3}$, where $b_3$ is the third Betti number
 of the compact space. Our ansatz is chosen so that the 3-form
 fluxes do not simultaneously have legs on both $A_3$ and $B_3$, so
 that $b_3$ is effectively $2$.)

 Let us now examine each of the two simple ways to cancel the $C_4$
 tadpole. One solution, which was considered in \cite{EvaModuli}, is to
 use $\overline{\mathrm{D3}}$-branes. This nice example, however, has an interesting
 instability. We
 know that in flat space, the gravitational attraction of the anti-branes
 is exactly cancelled by the electrical repulsion due to the R-R 5-form
 field strength. In the set-up here however, there is in fact a net attractive force between
 the anti-branes resulting from the polarization of the flux in
 between \cite{GiantInf}. The background $H_3 \wedge F_3$ can be thought of as an
 induced D3-brane charge. This charge screens the form-field
 repulsion, but since the graviational force is unchanged, there is a
 resultant attractive force. The analysis of \cite{GiantInf}
 indicates that this process of clumping of the anti-branes is
 very quick.

 The other option, using orientifold planes, is free
 of this instability, since by definition they cannot move. Therefore, we
 perform a stability analysis with O3-planes. We find,
 surprisingly, that the mass spectrum is in fact independent of both
 $g_s$ and $\alpha$, but depends only on the orientifold action. The
 orientifold action is $\Omega R (-1)^{F_L}$ \cite{Sen, FP, KST}, where
 $\Omega$ is the world-sheet parity, $R$ is the ${\bf Z}_2$
 action on $X_6$, and $F_L$ is the spacetime fermion number of the left-movers.
 We shall be looking at an ansatz with constant three-form fluxes
 $H_3$ and $F_3$ which are even under the ${\bf Z}_2$.
 The other fields such as the axion, dilaton, and the metric are also
 even. In this case, we have the result
 that most choices of $A_3$ and $B_3$ will lead to instabilities! The
 precise criterion we find is that there is an instability if there is
 an eigenfunction of the scalar Laplacian operator on $A_3$ (or $B_3$) which is {\it even} under the $\mathbf{Z}_2$
 action and has eigenvalue $3 \leq - R^2 \square < 22.1 $, where $R$ is the radius of curvature of $A_3$ (and $B_3$). The
 mode is tachyonic in the sense that its mass is below the Breitenlohner--Freedman bound
 \cite{BF}. Indeed, we give an explicit example where the compact space
 is chosen to be $S^3 \times S^3$, which turns out to be unstable.
 It would be interesting to have an explicit example of $A_3$ and
 $B_3$ with no eigenvalue of the Laplacian in the above range, or in
 which the orientifold action projects out the unstable mode.

  In this paper we shall carry out the stability analysis along the
 lines of \cite{Duff, Cosets, Biran, DFGHM}. In this
 approach, we shall be able to compute the spectrum of the entire KK tower of
 fluctuations about the background. Often the approach is to perform
 a zero-mode (or s-wave) reduction of the higher dimensional theory
 down to four dimensions. The advantage of the latter method is that one
 clearly sees the various contributions to the four-dimensional
 effective potential for the different fields such as the size of the
 compact space, dilaton, etc. Some of the contributions, like fluxes,
 contribute positively, while others, such as positive curvature of
 the compact space and negative tension objects, have negative
 contributions. The various `forces' stabilizing the moduli
 come at different orders in the string perturbation theory: for
example, the 4-d vacuum energy from compactifying on a compact
space and the energy from NS-NS fluxes are at string tree-level,
while examples of contributions at one-loop include R-R fluxes and
 gaugino condensates \cite{Heterotic}. In fact, sometimes the
 contribution could be even non-perturbative, for instance from
 Euclidean branes as in \cite{KKLT, KKLLMMT}. A nice review of these
 various types of contributions are listed in the lecture notes
 \cite{EvaModuli}. The term ``moduli stabilization'' is used to
 indicate solutions in which the potential has at least one minimum
 with finite non-zero values of every field. From the ten-dimensional
 point of view, this simply means that there are non-trivial solutions
 to the equations of motion.

 Although this s-wave effective
 potential is an excellent guide in computing the background, we should remind ourselves that it can
 be quite misleading in informing us of the stability of the
 background. In other words, the fact that the background sits at the
 minimum of the effective potential is not a guarantee for stability. In fact,
 there could be a tachyonic instability in a
 low-lying KK fluctuation which is not an s-wave.\footnote{See for instance
 the result in \cite{DFGHM} in which the lowest mass mode in the
 coupled scalar sector was a mode whose ``orbital angular momentum'' was
 ${3 \over 4} (q-1)^2$, where $q$ is the dimension of the compact manifold.} It is for this reason that we inspect the
 low-lying tower of states for this class of compactifications. It is a highly non-trivial matter to sieve
  through all the modes, and in practice one uses some judgement to
  pick out modes which are usually least massive to determine
  stability. Keeping this in perspective, we shall use the lessons
 learned in \cite{DFGHM, Shiromizu} (see also \cite{Campbell}) about product-space
 compactifications.

The most obvious source of instability for product spaces $X_q = A_n
 \times B_{q-n}$ is an instability in which one part, say $A_n$,
 expands, while the other part $B_{q-n}$ shrinks, preserving the total
 volume locally. In fact \cite{DFGHM, Shiromizu} showed that $(p+q)$-dimensional Einstein--Maxwell theories
 with a $q$-form flux, when compactified on a $q$-dimensional positively curved Einstein space, had only precisely this
instability. But results of that analysis do not apply to the
example that we have in mind because here we do not have a 6-form
flux threading the $A_3 \times B_3$. Instead, we have both R-R and
NS-NS 3-form fluxes threading each three-cycle. The presence of
the 3-form fluxes would lead one to expect that this mode would be
stabilized. More generally, the class of modes that we consider is
the coupled scalar sector, consisting of the traces of the metric
fluctuations on AdS, on
 $A_3$ and $B_3$,
the fluctuations of the dilaton and axion, and the fluctuation of
the R-R and NS-NS 3-form fields. We compute the masses of these modes.

  This paper is organized as follows. In
 Section \ref{Realistic} we explain the set-up and determine the background
 from a ten-dimensional perspective.\footnote{For a four-dimensional
 treatment of the background, see \cite{EvaModuli}.} We point out the
 two possible choices of cancelling the $C_4$ tadpole, by adding
 mobile $\overline {\mathrm{D3}}$-branes or by adding O3-planes, and then
 discuss possible instabilities. Section \ref{Vacua} discusses the
 growth of the number of $AdS_4$ vacua with 3-form fluxes. Finally,
 Section \ref{Detailed}, which is the most important section of the paper,
 determines the set of vacua which are stable.

\section{$\text{AdS}_{\bf 4}$ Vacua in IIB}
\label{Realistic}

The bosonic part of the IIB action in Einstein frame is
\cite{Polbook}
 \eqn{IIBAction}{\eqalign{
 S_{\text{IIB}} = {1 \over {2 \kappa_{10}^2}} \int & d^{10}x \sqrt{-g}
   \Big[R - {1 \over 2} (\partial \phi)^2 - {1 \over {2}} e^{2\phi}(\partial C_0)^2
       -{1 \over {2}}(C_0^2 e^{\phi}+ e^{- \phi})|H_{3}|^2 \cr &+C_0e^{\phi}H_3\cdot F_3
       - {1 \over 2} e^{\phi}|F_3|^2-{1 \over 4} |\tilde{F}_5|^2 \Big]
       +{1 \over {4\kappa_{10}^2}}\int{C_4\wedge F_3\wedge
       H_3} \,,}
  }
where $\tilde{F}_5$ is defined as
 \eqn{}{ \tilde{F}_5=dC_4-\frac12 C_2\wedge H_3+\frac12B_2\wedge
       F_3\,.
 }
The equations of motion obtained by varying this action are
 \eqn{EOM}{\eqalign{ R_{MN} &= {1 \over 2}
\partial_M \phi
\partial_N \phi + {1 \over 2} e^{2 \phi} \partial_M C_0 \partial_N C_0
+ (C_0^2 e^{\phi} + e^{-\phi}) \left[{1 \over {2 \cdot
2!}}H_{MP_1P_2}H_{N}^{\;\;P_1P_2} - {1 \over
    8}g_{MN}|H_3|^2 \right]
\cr  &- C_0 e^{\phi} \left[{1 \over {2 \cdot 2!}}
  H_{MP_1P_2} F_{N}^{\;\;P_1P_2} + {1 \over {2 \cdot 2!}}
  F_{MP_1P_2} H_{N}^{\;\;P_1P_2} - {1 \over 4}g_{MN}H_3 \cdot F_3 \right]
  \cr &+  e^{\phi} \left[{1 \over {2 \cdot 2!}}F_{MP_1P_2}F_{N}^{\;\;P_1P_2}
- {1 \over
    8}g_{MN}|F_3|^2 \right] +\frac{1}{4\cdot 4!}
  \tilde{F}_{MP_1P_2P_3P_4}\tilde{F}_N^{\;\;P_1P_2P_3P_4}\,, \cr 0 &= \square\phi-e^{2\phi}(\partial
C_0)^2-\frac12e^{\phi}|F_3|^2+C_0e^{\phi}H_3 \cdot F_3
-\frac12(C_0^2e^{\phi}-e^{-\phi})|H_3|^2 \,, \cr 0 &= \square
C_0+2\partial_{M}\phi\partial^{M}
 C_0-C_0e^{-\phi}|H_3|^2+e^{-\phi}H_3\cdot F_3 \,, \cr
0 &=
d*(C_0^2 e^{\phi}+e^{-\phi})H_3-d*(C_0e^{\phi} F_3)-F_3\wedge*\tilde{F}_5
\,, \cr 0 &=
d*(e^{\phi}F_3)-d*(C_0e^{\phi}H_3)+H_3\wedge*\tilde{F}_5 \,, \cr
0 &=d*\tilde{F}_5 - H_3\wedge F_3 \,, \quad \quad \quad
 \tilde{F}_5=*\tilde{F}_5\,.
 }}

We shall consider Freund--Rubin compactifications of the form
\eqn{ProdSpace}{\eqalign{
 ds^2 = & ds_{\text{AdS}_4}^2 + ds_{A_3}^2 + ds_{B_3}^2 \,, \cr
  H_{3} = & N_1\vol_{A_3}+N_2\vol_{B_3} \,, \quad
 F_{3} = Q_1\vol_{A_3}+Q_2\vol_{B_3}\,,
 }}
where $\vol_{A_3}$ is the volume form of $A_3$ and similarly for
 $B_3$. We use upper-case Roman letters $M, N, \ldots$, for indices on the
full $10$-dimensional spacetime, Greek letters $\mu, \nu, \ldots$,
for indices on AdS, lower-case Roman letters $a,b,\ldots$, for
indices on $A_3$, and $i,j, \ldots$, for indices on $B_3$. Since
we are dealing with Einstein spaces, the Ricci tensor is related
to the metric tensor as follows:
 \eqn{BackgroundAdS}{\eqalign{
 R_{\mu \nu} = -{3 \over L^2}g_{\mu \nu}\,, \qquad
 R_{a b} = {2 \over R_1^2}g_{ab}\,, \qquad
 R_{ij} = {2 \over R_2^2}g_{ij} \,,
 }}
where $L,R_1$, and $R_2$ are radii of curvature of $\text{AdS}_4$,
$A_3$, and $B_3$, respectively.

The IIB action has an SL(2,$\mathbb{Z}$) symmetry of the
axion-dilaton. We focus on the simplest ansatz and choose a zero
axion and constant dilaton. The condition $C_0 = 0$ implies
\eqn{axiontadpole}{N_1Q_1+N_2Q_2=0\,.
 }
A convenient parametrization of the four fluxes satisfying the
constraint (\ref{axiontadpole}) is
 \begin{equation}\label{fluxpara}
 N_1=\alpha N\,,\quad N_2=N\,,\quad Q_1=-\tfrac{1}{\alpha}Q\,,\quad Q_2=Q\,,
 \end{equation}
where we restrict $\alpha$, $N$, and $Q$ to taking on positive
values. Furthermore, requiring a constant dilaton results in
\eqn{dilback}{
 e^{\phi} = {{\alpha N} \over Q} \,.
}

Integrating the ${\tilde F_5}$ equation of motion over the compact space gives us a
 non-trivial Gauss's law constraint on the compact space $X_6$:
 $$0=\int_{X_6}H_3\wedge
F_3=(N_1Q_2-N_2Q_1)\cdot(\text{volume of } X_6),$$ which is
clearly incompatible with (\ref{axiontadpole}). To make the net
three-brane charge vanish we introduce local objects which carry
this charge such as D3 (and $\overline{\mathrm{D3}}$) branes and
O3 (and $\overline{\mathrm{O3}}$) planes as in \cite{GKP}. These
sources are localized on the compact space and extend along AdS.
The full action including the localized sources is now
\eqn{addlocal}{\eqalign{ S = S_{\text{IIB}} + S_{{\rm local}}
\quad \quad
 {\rm where} \quad \quad S_{{\rm local}} = - T_3 \int d^4 x \sqrt{-g} + \mu_3 \int
 C_4 \,.
}} $T_3$ and $\mu_3$ are the three-brane tension and charge,
respectively. This modifies the $\tilde{F}_5$ equation of motion
(\ref{EOM}) to
 \eqn{f5eommod}{d*\tilde{F}_5=H_3\wedge
 F_3+2\kappa_{10}^2 T_3 \left(\rho_D - {1 \over 4} \rho_O \right) \,,
 }
where $\rho_D$ and $\rho_O$ are the number densities of D3-branes
 and O3-planes on the compact space, respectively. The
corresponding global constraint is now \cite{EvaModuli}
 \eqn{tadpole}{\frac{1}{2 (2\pi)^4(\alpha')^2}\int_{X_6}H_3\wedge F_3=N_{\overline{D3}}-N_{D3}
 + {1 \over 4} \left(N_{O3} -
 N_{\overline{O3}} \right)\,.
 }
Note that we have put an extra factor of $2$ on the left-hand
side, anticipating $N_{O3}\neq 0$. This arises in the presence of
O3-planes because in (\ref{tadpole}) we are
 integrating over the full covering space instead of the ${\bf Z}_2$
 quotient.

 Let us recall that the 3-form fluxes are quantized; that is,
 \eqn{fluxquant}{\eqalign{
 n_1&=\frac{1}{(2\pi)^2\alpha'}\int_{A_3}H_3\,,\qquad
 n_2 =\frac{1}{(2\pi)^2\alpha'}\int_{B_3}H_3\,,\cr
 q_1&=\frac{1}{(2\pi)^2\alpha'}\int_{A_3}F_3\,,\qquad
 q_2 =\frac{1}{(2\pi)^2\alpha'}\int_{B_3}F_3\,,
 }}
 are all integers. In the presence of orientifold planes,
 because of the ${\bf Z}_2$ action on $A_3$ and $B_3$, we restrict the
 integers $n_1,n_2,q_1,q_2$ to even values.\footnote{See however,
 \cite{FP, KST} for how this restriction could be lifted by
 considering O3-planes with discrete fluxes and with $+1/4$ D3 charge.}
 Furthermore, from Eq.~(\ref{fluxpara}) we obtain $\alpha = n_1/n_2 =
 |q_2/q_1|$, which prevents $\alpha$ from varying continuously and
 restricts it to taking only rational values. In fact, there is a
 further restriction that $\alpha$ has to satisfy, which comes from
 the tadpole constraint. To see this, let us combine \fluxquant\
 with (\ref{tadpole}), which allows us to express the number
 of localized objects in terms of three-form flux quanta:
\eqn{LocalNo}{
 {1 \over 2}(n_1 q_2  - n_2 q_1) = {1 \over 2}\Big(\alpha + {1 \over \alpha}\Big)nq
 = N_{\overline{D3}}-N_{D3} + {1 \over 4} \left(N_{O3} -
 N_{\overline{O3}} \right) \,,
} where $n = n_2$ and $q=q_2$. Since we restrict ourselves to the
cases with positive $\alpha$, $n$, and $q$, we shall now consider
the following two cases: (a) only $N_{\overline{D3}} \neq 0$, and
(b) only $N_{O3} \neq 0$.

In the first option, one puts mobile anti-D3 branes with their
world-volumes along $AdS_4$ and each localized as a point on
$X_6$. Let us examine what happens to the system if we start with
a uniform distribution of branes on $X_6$. To determine whether
this system is stable, it suffices to consider the interaction of
two anti-branes. In flat space, a pair of anti-D3 branes would be
mutually BPS --- their attractive gravitational force exactly
cancels the repulsive R-R force. When $A_3$ and $B_3$ are
three-spheres, this is also true to leading order. But in our
set-up there is more to the picture. There is also a net $H_3
\wedge F_3$ flux threading the three-spheres. From (\ref{tadpole})
it is clear that the flux acts as a cloud of positive 3-brane
charge. This screens the negative charge on each anti-brane and
therefore, the gravitational attraction is now stronger than the
R-R repulsive force. This interesting instability due to flux
polarization was encountered in \cite{GiantInf}. To sum up, a
Jeans instability sets in and the anti-branes clump on the compact
space, possibly followed by the interesting dynamics studied in
great detail in \cite{GiantInf}. It would be interesting to
investigate whether supersymmetry is restored in the end, but we
shall not do so here.

 The second option, that of introducing O3-planes, is what we shall
 look into in some detail in this paper. Like the anti-D3 branes
 above, these orientifold planes also lie completely along AdS. It is
 important to realize that the number of such planes that can be
 introduced is not arbitrary, but depends on the symmetries of the
 compact space. More specifically, one looks for fixed points of the
 action of a subgroup $\Gamma$ of the isometry group of $X_6$. For
 the example in which both $A_3$ and $B_3$ are simply three-spheres,
 $\Gamma$ is a subgroup of $\mathrm{SO}(4) \times
 \mathrm{SO}(4)$. The number of such orientifold planes which are allowed are
 determined by the choice of $X_6$ and $\Gamma$. We shall denote the number of such O3-planes
 by $t$ in this paper, and view (\ref{LocalNo}) as a constraint on
 the fluxes. Combining this with Eqs.~(\ref{fluxpara}) and (\ref{dilback}), we can
 now parameterize the space of vacua with three numbers $(g_s, \alpha, t)$, where $g_s =
 \alpha N/Q$ is determined from the background value of the dilaton,
 $\alpha$ can be thought of as an asymmetry parameter measuring the
 ratio of a given kind of flux on $A_3$ to that along $B_3$, and
 finally, $t=N_{O3}$ is just the number of O3-planes.

The introduction of localized objects changes the stress-energy
tensor $T_{MN}$ and so the background geometry gets corrected. To
compute the radii of curvature $L$, $R_1$, and $R_2$, let us write
Einstein's equation as
\eqn{metriceom}{R_{MN}=\kappa_{10}^2(T_{MN}-\frac18g_{MN}T^P_P)\,.
 }
Let us denote the collective coordinates on $A_3$ by $y$ and those
on $B_3$ by $z$, and let $y_k$ and $z_k$ be the coordinates of the
$k$-th orientifold plane on $X_6$. Then their contribution to the
stress-energy tensor is given by
 \eqn{tmnloc}{T_{\mu\nu}^{\text{loc}}={T_3 \over 4} g_{\mu\nu}\sum_{k}\delta^3(y-y_k)\delta^3(z-z_k),\quad
 T_{ab}^{\text{loc}}=T_{ij}^{\text{loc}}=0\,,
 }
where $-T_3/4$ is the tension of an O3-plane (so that $2
\kappa_{10}^2
 T_3 = (2 \pi)^4 \alpha'^2$). Together with the
contribution from the supergravity fields in (\ref{EOM}),
Einstein's equations (\ref{metriceom}) now reads
 \eqn{Einstein1}{R_{\mu\nu}=
 -\frac{3}{L^2}g_{\mu\nu}=-\frac14g_{\mu\nu}\Big(\alpha + {1 \over \alpha}\Big)NQ
 +\frac18\kappa_{10}^2T_3g_{\mu\nu}\sum_{k}\delta^3(y-y_k)\delta^3(z-z^k)
 }
along the AdS directions. Taking traces of $R_{MN}$ first along
$A_3$ and then along $B_3$ we also get
 \eqn{Einstein2}{\eqalign{
 R^a_{\;a}&=\frac{6}{R_1^2}=\frac34\Big(\alpha + {1 \over \alpha}\Big)NQ-\frac38\kappa_{10}^2T_3\sum_{k}\delta^3(y-y_k)\delta^3(z-z_k)\,,\cr
 R^i_{\;i}&=\frac{6}{R_2^2}=\frac34\Big(\alpha + {1 \over \alpha}\Big)NQ-\frac38\kappa_{10}^2T_3\sum_{k}\delta^3(y-y_k)\delta^3(z-z_k)\,.
 }}
From now on, we denote the (common) radius of curvature of $A_3$
and $B_3$ by $R$.

In order for the supergravity approximation to hold, the radius of
curvature of our background must be large compared to string
scale. Therefore, we take the limit $R/\sqrt{\alpha'}\gg 1$ with
dimensionless combinations $RN$ and $RQ$ fixed. From the
quantization condition (\ref{fluxquant}), we see that this is the
limit of large number of flux quanta on $A_3$ and $B_3$, and hence
we need a large number of O3-planes. Now, the number of O3-planes
per unit volume of $X_6$ scales like $nq/R^6\sim
1/R^2(\alpha')^2$, so the distance between adjacent O3's will
roughly be $\sim (R\alpha')^{1/3}$. Therefore, in the low energy
limit where we are only interested in the length scale, say
$\lambda>R^{2/3}(\alpha')^{1/6}$, we will not be able to see
individual O3-planes. Rather, in this low energy limit, we can
replace their delta function distribution in (\ref{tmnloc}) by a
uniform distribution \cite{GKP,GP,SSG} with density
 \eqn{rhoapprox}{\rho= {2 \over {(2\pi)^4 \alpha'^2}}\Big(\alpha + {1 \over
 \alpha}\Big)NQ\,.
 }
Then $\tilde{F}_5=0$ is a solution of (\ref{f5eommod}) in this
limit, and Einstein's equations (\ref{Einstein1}) and
(\ref{Einstein2}) determine the radii of AdS and those of $A_3$
and $B_3$ in terms of 3-form fluxes. Our background is:
 \eqn{adsradius}{\frac{3}{L^2}=\frac{2}{R_1^2}=\frac{2}{R_2^2} ={1 \over 8} \Big(\alpha + {1
     \over \alpha}\Big) NQ\,, \quad \quad e^{\phi} = \frac{\alpha N}{Q} \,.}

\section{Number of Vacua and the 4D Viewpoint}
\label{Vacua}

 The first thing one wants to know about the holographic dual of a string
 theory in AdS is the
 number of degrees of freedom of the field theory. For instance, for
 string theory on $AdS_5 \times S^5$, which is dual to
 ${\cal N} =4$ theory with gauge group ${\mathrm{SU}(N)}$, has $N^2$ degrees
 of freedom. It turns out that AdS/CFT allows one to predict the
 degrees of freedom at strong coupling by the Susskind--Witten
 arguments \cite{SW}. On the gravity side one
 computes the cosmological constant of the reduced theory (in the
 $AdS_5 \times S^5$ example, this is the 5-d cosmological constant);
 the entropy is then inversely proportional to this.  For 4-d theories
 the entropy is
 \eqn{Entropy}{
 S \sim {1 \over {l_4^2 \Lambda_4^E}} \,,
} where $l_4$ is the 4-d Planck length and $\Lambda_4^E$ is the
4-d
 cosmological constant (measured in the Einstein frame). Let us compute
 the entropy for our $AdS_4$ vacua which result from $A_3 \times B_3$
 compactifications that we have considered in this paper.

We shall compute $\Lambda_4^E$ from the reduced 4-d effective
potential that we mentioned in the Introduction. This will tell us
about the shape of the s-wave effective potential, reminding us of
the various positive and negative contributions to the vacuum
energy. In this section we follow the treatment in
\cite{EvaModuli}.

 The string frame cosmological term is obtained by integrating various
 terms in the string-frame action over the compact space $X_6$. One gets
\eqn{4dLambda}{\eqalign{
  \Lambda_4^s &\sim e^{-2 \phi} \left( - {{R_1^3 R_2^3} \over R_1^2}  -{{R_1^3 R_2^3}
     \over R_2^2}  + {{n_1^2 R_2^3} \over R_1^3} + {{n_2^2 R_1^3}
     \over R_2^3} \right) + \left({{q_1^2 R_2^3} \over R_1^3} + {{q_2^2
     R_1^3} \over R_2^3}\right)  - t e^{-\phi}  \,,
 }}
where we have separated the contributions due to the NS-NS fields,
the R-R fields, and finally the O3-planes. Everything is measured
in string units. The metric in the string frame is related to that
in the
 Einstein frame by
 \eqn{Frame}{
 g^E_{MN} = e^{-\phi/2} g^s_{MN} \,.
} From this, one can show that the 4-d cosmological constant in
the Einstein frame and string frame are related by
 \eqn{frameconv}{
 l_4^2 \Lambda_4^E = {e^{4 \phi} \over V_X^2} \Lambda_4^s \,,
 }
where $V_X$ is the volume of the compact space. To get just the
scalings, we assume $n_1=n_2 =n$ and $q_1=q_2 =q$ in this section
(in the notation of the rest of the paper, this is the same as
setting $\alpha =1$). Minimizing the energy (\ref{4dLambda}) with
respect to $R_1$ and $R_2$ immediately yields $R_1 = R_2 \sim
{\sqrt n}$, while minimizing with respect to $\phi$ yields
$e^{\phi} \sim n/q$. These are exactly the same relations as we
obtained in (\ref{adsradius}) by solving the ten-dimensional
equations of motion.\footnote{To show this recall that $n \sim
NR_E^3$,
 and $q \sim Q R_E^3$. From (\ref{Frame}) one has $R_E^2 = e^{-\phi/2}
 R_s^2$.} Inserting these in (\ref{4dLambda}), we find that
each term scales the same, and that
 \eqn{4dentropy}{
 {1 \over S} \sim \left({n \over q}\right)^4 {1 \over {n^3 n^3}}
 \left[{{n^{3/2} n^{3/2}} \over n} {q^2 \over n^2} + \cdots \right]
 \sim {1 \over (qn)^2} \sim {1 \over t^2}\,.
} So we see that the entropy of these solutions depends on the
number of orientifold planes (or $\overline{\mathrm{D3}}$-branes
for that matter) as:
 \eqn{Strel}{
 S \sim t^2 \,.
 }

It is interesting to see how this compares with the number of flux
vacua obtained by the Bousso--Polchinski arguments
\cite{BoussoPol}. In \cite{EvaModuli} for instance, this was
heuristically sketched out and it was found that
 \eqn{VacNum}{
 N_{\rm{Vacua}} \sim {t^{b_3} \over b_3!} \,,
} where $b_3$ is the third Betti number. For $X_6 = S^3 \times
S^3$, we find that (\ref{Strel}) and (\ref{VacNum}) agree. For
other compact spaces $b_3$ may be different, but the fact that we
do not allow the 3-form fluxes to simultaneously have legs on both
$A_3$ and $B_3$ is responsible for the dependence above. It is
very interesting to ask if one can account for the entropy in
(\ref{Strel}) by moving into the Coulomb branch and counting
string junctions as in \cite{Junction}.

\section{How Many of These Vacua Are Stable?}
\label{Detailed}

In anti-de Sitter space, the criterion for stability was computed
by Breitenlohner and Freedman \cite{BF}, who showed that in
$\mathrm{AdS}_4$, for a (normalizable) scalar field to be stable,
its mass has to be greater than the so-called BF bound:
\eqn{Bound}{
 m^2 L^2 \geq - {9 \over 4} \,.
 }

In this section, we shall examine perturbative stability of the
vacuum solution described in Section \ref{Realistic} against small
fluctuations. The particular excitations we shall be interested in
are fields which are scalars in the dimensionally reduced theory
on $AdS_4$. The ten-dimensional origin of these scalars are the
trace of the graviton on the $A_3 \times B_3$ (or alternatively,
as we shall see, the trace on the metric fluctuations of $AdS_4$),
the dilaton, the axion, and scalar fluctuations of the 3-form
fields.

\subsection{Fluctuations}\label{fluctuations}

Let us introduce the following notations for the fluctuation of
metric:
\begin{gather}
 \delta g_{\mu \nu} = h_{\mu \nu} = H_{\mu \nu} - {1 \over 2} g_{\mu \nu} (h^a_a+h^i_i)
 \,, \label{MetricAdS}
 \\
  \delta g_{\mu a} = h_{\mu a}, \qquad
 \delta g_{\mu i} = h_{\mu i},  \qquad
 \delta g_{ab} = h_{ab}, \qquad
 \delta g_{ij} = h_{ij}, \qquad
 \delta g_{ai} = h_{ai}. \label{Metricint}
 \end{gather} In Eq.~(\ref{MetricAdS}), we have
defined the standard linearized Weyl shift on $h_{\mu \nu}$. It
will be useful to decompose $H_{\mu \nu}$, $h_{ab}$, and $h_{ij}$
into trace and traceless parts:
\begin{eqnarray}\label{MetricAdSint}
 H_{\mu \nu} = H_{(\mu \nu)} + {1 \over 4} g_{\mu \nu} H^\rho_\rho \,,
 \quad h_{ab} = h_{(ab)} + {1 \over 3}
 g_{ab} h^c_c \,, \quad h_{ij}=h_{(ij)}+\frac13g_{ij}h^k_k,
\end{eqnarray}
so that $g^{\mu \nu} H_{(\mu \nu)} = g^{ab} h_{(ab)}= g^{ij}
h_{(ij)} = 0$. We also introduce the fluctuation of the form
fields\footnote{Beware that we use the same letter $h$ to denote
both the metric and NS-NS 3-form perturbations; they are
distinguished by the number of indices they carry.}
\begin{equation}\label{formfluc}
 \delta H_3 \equiv h_3 = db_2 \qquad
 \delta F_3 \equiv f_3 = dc_2\,,
\end{equation}
and the fluctuations $\delta\phi$ and $\delta C_0$ of dilaton and
axion fields.

To fix the internal diffeomorphisms and gauge freedom, and to
simplify our analysis, we impose the de Donder--type gauge
conditions
\begin{eqnarray}
\label{DDGauge} \nabla^a h_{(ab)} = \nabla^i h_{(ij)} = 0 \,,
\quad \nabla^a h_{\mu a}=\nabla^i h_{\mu i}=\nabla^a
h_{ai}=\nabla^i h_{ai} = 0 \,,
\end{eqnarray}
on the metric fluctuations and the Lorenz-type conditions
\begin{equation}\label{LorentzGauge}
\begin{split}
\nabla^a b_{\mu a}& =\nabla^a b_{ai}=\nabla^a b_{ab}=0\,, \quad
\nabla^i b_{\mu i} =\nabla^i b_{ai}=\nabla^i b_{ij}=0\,, \\
\nabla^a c_{\mu a}& =\nabla^a c_{ai}=\nabla^a c_{ab}=0\,, \quad
\nabla^i c_{\mu i} =\nabla^i c_{ai}=\nabla^i c_{ij}=0\,,
\end{split}
\end{equation}
on the form fluctuations. As we mentioned above, we restrict our
analysis to scalar modes in AdS. In the above gauge, we can expand
the fluctuations in terms of scalar harmonics of $A_3$ and $B_3$
as follows:
 \begin{align}
 \delta\phi &= \sum_{IJ} \delta\phi^{IJ}(x)
Y^I(y) Z^J(z) \,,  &\delta C_0 &= \sum_{IJ} \delta C_0^{IJ}(x)
Y^I(y) Z^J(z) \,, \label{Taufluct}\\
 b_{ab} &=  \sum_{IJ}
b_{(1)}^{IJ}(x) \epsilon_{ab}^{\;\;\; c}\nabla_cY^I(y) Z^J(z) \,,
& b_{ij} &= \sum_{IJ} b_{(2)}^{IJ}(x) Y^I(y)
\epsilon_{ij}^{\;\;\; k}\nabla_k Z^J(z) \,, \label{b2fluct} \\
 c_{ab} &=
\sum_{IJ} c_{(1)}^{IJ}(x) \epsilon_{ab}^{\;\;\;
c}\nabla_cY^I(y)Z^J(z) \,, & c_{ij} &=  \sum_{IJ} c_{(2)}^{IJ}(x)
Y^I(y) \epsilon_{ij}^{\;\;\; k}\nabla_k Z^J(z) \,,
\label{c2fluct} \\
  h^a_a &= \sum_{IJ} \pi_{(1)}^{IJ}(x) Y^I(y)
Z^J(z) \,, & h^i_i &= \sum_{IJ} \pi_{(2)}^{IJ}(x) Y^I(y) Z^J(z)
\,, \label{internalfluct}\\
 H^\mu_\mu & = \sum_{IJ} H^{IJ}(x)
Y^I(y) Z^J(z) \,. & &  \label{adsfluct}
 \end{align}
Here, $Y^I$ and $Z^J$ are the scalar spherical harmonics on $A_3$ and
$B_3$, respectively. Note that the gauge
conditions (\ref{LorentzGauge}) imply that the two-forms
$b_{ab}$, $b_{ij}$, $c_{ab}$, $c_{ij}$ are co-exact on the
respective 3-spaces; in (\ref{b2fluct}) and (\ref{c2fluct}), we used the fact that
co-exact two-forms on 3-manifolds can be expressed as the Hodge dual
of the exterior derivatives of scalars.

\subsection{Fluctuation Equations}
\label{EquationSec}

In this section we derive the linearized equations of motion for
the fluctuations defined in the previous subsection. We shall
follow the analysis of \cite{DFGHM} quite closely. Although
product spaces were dealt with in that paper, the set-up differed
in that there was a 6-form field threading the $S^3 \times S^3$,
instead of 3-form fields like we have here.

Let us consider the dilaton equation first. Expanding the dilaton
equation of motion to first order in fluctuations yields
\eqn{flucdilaton}{\eqalign{
 0 = \square\delta\phi &- \frac12 (e^{-\phi}|H_3|^2+e^{\phi}|F_3|^2)\delta \phi \cr &+ {e^{-\phi} \over 2} \left[ {2
    \over {3!}} H_{abc} h^{abc} + {2 \over {3!}} H_{ijk} h^{ijk} - (N_1^2h^a_{a}
    +N_2^2 h^i_i) \right] \cr
    & - {e^{\phi} \over 2} \left[ {2
    \over {3!}} F_{abc} f^{abc} + {2 \over {3!}} F_{ijk} f^{ijk} - (Q_1^2h^a_{a}
    +Q_2^2  h^i_i) \right] \,.
  }}
As introduced in Section \ref{fluctuations}, we write the
fluctuations of the field strengths as $h_3=db_2$ and $f_3=dc_2$,
where
 \eqn{potfluc}{\eqalign{
  b_2 & = \frac12 b_{ab} dy^a\wedge dy^b + \frac12 b_{ij} dz^i \wedge dz^j\,, \cr
  c_2 & = \frac12 c_{ab} dy^a\wedge dy^b +
      \frac12 c_{ij} dz^i \wedge dz^j\,.
  }}
Note that we have set the fluctuation modes $b_{\mu\nu}$, $b_{\mu
a}$, $b_{\mu i}$, $c_{\mu\nu}$, $c_{\mu a}$, $c_{\mu i}$ to zero,
because we are interested only in the scalar fluctuations. It can
be readily shown that these modes do not enter in the equations of
motion for the scalar modes in AdS. Then we can expand each mode
in terms of the spherical harmonics as in (\ref{b2fluct}) and
(\ref{c2fluct}). After a straightforward algebra, the dilaton
fluctuation equation (\ref{flucdilaton}) reduces to\footnote{We
use the following notation: $\square_x \equiv g^{\mu\nu}
\nabla_\mu \nabla_\nu$, $\square_y \equiv g^{ab} \nabla_a
\nabla_b$, and $\square_z \equiv g^{ij} \nabla_i \nabla_j$.}
 \eqn{finaldilaton}{\eqalign{
 \bigg[ & \left\{  \square_x + \square_y +\square_z
 -\Big(\alpha+\frac{1}{\alpha}\Big)NQ \right\}\delta\phi^{IJ}-\frac12\Big(\alpha-\frac{1}{\alpha}\Big) NQ(\pi_{(1)}^{IJ}
 -\pi_{(2)}^{IJ})  \cr
 & + {Q \over {\alpha N}} \left( N_1 \square_y b_{(1)}^{IJ} + N_2 \square_z b_{(2)}^{IJ}\right)
 - {{\alpha N} \over Q} \left( Q_1 \square_y c_{(1)}^{IJ}
   + Q_2 \square_z c_{(2)}^{IJ} \right) \bigg]Y^IZ^J=0 \,.
 }}

A similar calculation results in the following fluctuation
equation for the axion:
 \eqn{finalaxion}{\eqalign{
 \bigg[ & \left\{  \square_x + \square_y +\square_z
 -\Big(\alpha+\frac{1}{\alpha}\Big)NQ \right\}\delta C_0^{IJ}+ {Q^2 \over \alpha}(\pi_{(1)}^{IJ}
 -\pi_{(2)}^{IJ}) \cr
 & +\frac{Q}{\alpha N} \left(Q_1 \square_y b_{(1)}^{IJ} + Q_2 \square_z b_{(2)}^{IJ}
 + N_1 \square_y c_{(1)}^{IJ}
   + N_2 \square_z c_{(2)}^{IJ} \right)  \bigg]Y^IZ^J=0 \,.
 }}

The fluctuation equations for $H_3$ and $F_3$ are similar, so we
shall elaborate the computation for the $H_3$ fluctuation and
write down only the answer for the $F_3$ fluctuation. Using the
fact that $C_0=0$ in our background, the variation of $H_3$
equation of motion reads
 \eqn{FormFluct}{\eqalign{ d
 (\delta *)
 (e^{-\phi} H_3) - d * (e^{-\phi} \delta \phi H_3) + d * (e^{-\phi}
 h_3) -d * (\delta C_0 e^{\phi}F_3) = 0 \,. }
 }
Here, $\delta *$ in the first term denotes the variation of the
Hodge star operator due to the metric fluctuation, which turns out
to be
 \eqn{starvar}{\eqalign{
  d(\delta
  *)(e^{-\phi}H_3)=\frac{Q}{\alpha N} \bigg[ & \frac{N_1}{2}d(h^{\mu}_{\mu}-h^a_a
  +h^i_i)\wedge\vol_{AdS_4}\wedge\vol_{B_3} \cr - & \frac{N_2}{2}d(h^{\mu}_{\mu}+h^a_a-h^i_i)\wedge
  \vol_{AdS_4}\wedge\vol_{A_3}\bigg].
  }}
Then we can again write $h_3=db_2$, and expand the fluctuation
fields in terms of spherical harmonics on $A_3$ and $B_3$. This
gives rise to the following set of equations:
 \begin{equation}\begin{split}
 d\Big[ &\Big\{\frac{N_1}{2}(H^{IJ}-3\pi^{IJ}_{(1)}-\pi^{IJ}_{(2)})-N_1\delta\phi^{IJ}
  -\frac{\alpha^2N^2}{Q^2}Q_1\delta C_0^{IJ} \\ &
 +\left(\square_x+\square_y+\square_z\right)b^{IJ}_{(1)}\Big\}Y^IZ^J\Big]\wedge\vol_{AdS_4}\wedge\vol_{B_3}=0\,,
 \label{finalh31} \end{split}\end{equation}

  \begin{equation}\begin{split}
 d\Big[&\Big\{\frac{N_2}{2}(H^{IJ}-\pi^{IJ}_{(1)}-3\pi^{IJ}_{(2)})-N_2\delta\phi^{IJ}-\frac{\alpha^2N^2}{Q^2}Q_2\delta
 C_0^{IJ}\\ &
 +\left(\square_x+\square_y+\square_z\right)b^{IJ}_{(2)}\Big\}Y^IZ^J\Big]\wedge\vol_{AdS_4}\wedge\vol_{A_3}=0\,.
 \label{finalh32}
 \end{split}\end{equation}
Similarly one can derive the fluctuation equations for $F_3$:
 \begin{equation}\begin{split}
 d\Big[&\Big\{\frac{Q_1}{2}(H^{IJ}-3\pi^{IJ}_{(1)}-\pi^{IJ}_{(2)})+Q_1\delta\phi^{IJ}-N_1\delta
 C_0^{IJ} \\
 & +\left(\square_x+\square_y+\square_z\right)c^{IJ}_{(1)}\Big\}Y^IZ^J\Big]\wedge\vol_{AdS_4}\wedge\vol_{B_3}=0\,,
 \label{finalf31}
 \end{split}\end{equation}
 \begin{equation}\begin{split}
 d\Big[&\Big\{\frac{Q_2}{2}(H^{IJ}-\pi^{IJ}_{(1)}-3\pi^{IJ}_{(2)})+Q_2\delta\phi^{IJ}-N_2\delta
 C_0^{IJ} \\ &
 +\left(\square_x+\square_y+\square_z\right)c^{IJ}_{(2)}\Big\}Y^IZ^J\Big]\wedge\vol_{AdS_4}\wedge\vol_{A_3}=0\,.
 \label{finalf32}
 \end{split}\end{equation}

Let us now consider Einstein's equations to linear order in
fluctuations. We shall need to expand the Ricci tensor to linear
order; in our conventions, this is:
 \eqn{RicciExpand}{\eqalign{
  \delta R_{MN} &= -{1 \over 2}[(\square_x + \square_y + \square_z)h_{MN} + \nabla_M \nabla_N h_{P}^{P}
  - \nabla_M \nabla^P h_{PN} - \nabla_N \nabla^P h_{PM} \cr
 &\qquad{}- 2 R_{MPQN} h^{PQ} - R_{M}^{\;\; P}h_{NP} - R_{N}^{\;\; P} h_{MP}] \,.
 }}
In terms of the graviton modes defined in
Eqs.~(\ref{MetricAdS})--(\ref{MetricAdSint}), the components of
$\delta R_{MN}$ with both indices in the internal manifold are
written as
 \begin{equation}\begin{split}
 \delta R_{ab} =  -\tfrac12 & [(\square_x+\square_y+\square_z)
 h_{(ab)}-2 R_{acdb}h^{(cd)}-R_a^{\;\; c}h_{(bc)}-R_b^{\;\;
 c}h_{(ac)} \\
   & +\tfrac13   g_{ab}(\square_x+\square_y+\square_z) h^c_c + \nabla_a
 \nabla_b (H^{\mu}_{\mu}-\tfrac53 h^c_c -
 h^i_i) \\
 & - \nabla_a \nabla^{\mu}h_{\mu b}-
 \nabla_b \nabla^{\mu} h_{\mu a}]\,, \label{Ricci1}
 \end{split}\end{equation}
 \begin{equation}\begin{split}
 \delta R_{ij} = -\tfrac12 & [(\square_x+\square_y+\square_z)
 h_{(ij)}-2 R_{iklj}h^{(kl)}-R_i^{\;\; k}h_{(jk)}-R_j^{\;\;
 k}h_{(ik)} \\
  & +\tfrac13 g_{ij}(\square_x+\square_y+\square_z) h^k_k +
 \nabla_i
 \nabla_j (H^{\mu}_{\mu}-h^c_c-\tfrac53 h^k_k
 ) \\ & - \nabla_i \nabla^{\mu}h_{\mu j}-
 \nabla_j \nabla^{\mu} h_{\mu i}]\,. \label{Ricci2}
 \end{split}\end{equation}
We have used the gauge conditions (\ref{DDGauge}) to simplify some
of the terms in the above expressions.

On the other hand, the variation of the stress-energy tensor
$\bar{T}_{MN}=
 T_{MN}-\frac18 g_{MN}T^P_P$ is given as follows:
  \begin{equation}\begin{split}
 \delta\bar{T}_{ab} = e^{-\phi} & \bigg[-\Big(\frac{1}{2\cdot
 2!}H_{acd}H_b^{\;\; cd}-\frac{2}{2\cdot
 8}g_{ab}|H_3|^2\Big)\delta\phi \\ &
 +\frac{1}{2\cdot
 2!}(h_{acd}H_b^{\;\; cd}+H_{acd} h_b^{\;\; cd} - 2
 h^{cd}H_{ace}H_{bd}^{\;\;\;\; e}) - \frac{2}{2\cdot 8}h_{ab}|H_3|^2 \\ & - \frac{2}{2\cdot 8}
 g_{ab}\Big(\frac{2}{3!}H^{cde}h_{cde}+\frac{2}{3!}H^{ijk}h_{ijk}-(N_1^2h^c_c+N_2^2h^i_i)\Big)\bigg]\\
 +  e^{\phi} & \bigg[\Big(\frac{1}{2\cdot 2!}F_{acd}F_b^{\;\;
 cd}-\frac{2}{2\cdot 8}g_{ab}|F_3|^2\Big)\delta\phi \\ &
 +\frac{1}{2\cdot
 2!}(f_{acd}F_b^{\;\; cd}+F_{acd} f_b^{\;\; cd} - 2
 h^{cd}F_{ace}F_{bd}^{\;\;\;\; e}) - \frac{2}{2\cdot 8}h_{ab}|F_3|^2 \\ & - \frac{2}{2\cdot 8}
 g_{ab}\Big(\frac{2}{3!}F^{cde}f_{cde}+\frac{2}{3!}F^{ijk}f_{ijk}-(Q_1^2h^c_c+Q_2^2h^i_i)\Big)\bigg]\\
 - e^{\phi}& \Big[{1 \over {2 \cdot 2!}} H_{acd} F_b^{\;\;cd} + {1
 \over {2 \cdot 2!}} F_{acd} H_b^{\;\;cd}\Big] \delta C_0 +
 \kappa_{10}^2 \delta \bar{T}_{ab}^{\text{loc}} \,,
 \label{emtensor1}
 \end{split}\end{equation}
 \begin{equation}\begin{split}
 \delta\bar{T}_{ij} =  e^{-\phi} & \bigg[-\Big(\frac{1}{2\cdot
 2!}H_{ikl}H_j^{\;\; kl}-\frac{2}{2\cdot
 8}g_{ij}|H_3|^2\Big)\delta\phi \\ & + \frac{1}{2\cdot
 2!}(h_{ikl}H_j^{\;\; kl}+H_{ikl} h_j^{\;\; kl} - 2
 h^{kl}H_{ikm}H_{jl}^{\;\;\;\; m}) - \frac{2}{2\cdot 8}h_{ij}|H_3|^2 \\ & - \frac{2}{2\cdot 8}
 g_{ij}\Big(\frac{2}{3!}H^{abc}h_{abc}+\frac{2}{3!}H^{klm}h_{klm}-(N_1^2h^a_a+N_2^2h^k_k)\Big)\bigg]\\
 +  e^{\phi} & \bigg[ \Big(\frac{1}{2\cdot 2!}F_{ikl}F_j^{\;
 kl}-\frac{2}{2\cdot 8}g_{ij}|F_3|^2\Big)\delta\phi \\ &+
 \frac{1}{2\cdot
 2!}(f_{ikl}F_j^{\;\; kl}+F_{ikl} f_j^{\;\; kl} - 2
 h^{kl}F_{ikm}F_{jl}^{\;\;\;\; m}) - \frac{2}{2\cdot 8}h_{ij}|F_3|^2 \\ & - \frac{2}{2\cdot 8}
 g_{ij}\Big(\frac{2}{3!}F^{abc}f_{abc}+\frac{2}{3!}F^{klm}f_{klm}-(Q_1^2h^a_a+Q_2^2h^k_k)\Big)\bigg]\\
 - e^{\phi}& \Big[{1 \over {2 \cdot 2!}} H_{ikl} F_j^{\;\;kl} + {1
 \over {2 \cdot 2!}} F_{ikl} H_j^{\;\;kl}\Big] \delta C_0 +
 \kappa_{10}^2 \delta \bar{T}_{ij}^{\text{loc}} \,.
 \label{emtensor2}
 \end{split}\end{equation}
The contributions $\delta \bar{T}_{ab}^{\text{loc}}$ and $\delta
\bar{T}_{ij}^{\text{loc}}$ from O3-planes are calculated in the
following way. Recall from Eq.~(\ref{tmnloc}) that
 \eqn{LocFluc}{
 \kappa_{10}^2 \bar{T}_{ab}^{\text{loc}} = -{{ \kappa_{10}^2 T_3 \rho} \over
   8} g_{ab} \,,
} where $-T_3/4$ is the tension of an O3-plane, and $\rho$ is
their density. By considering the metric fluctuation in the
 internal space and the resulting fluctuation of $\rho$, we obtain
  \eqn{locfluc}{
 \kappa_{10}^2 g^{ij} \delta \bar{T}_{ij}^{\text{loc}} = {{NQ} \over {16}} \Big(\alpha
 + {1 \over \alpha}\Big) \left( \pi_1 + 3 \pi_2 \right) \,, \quad
 \kappa_{10}^2 g^{ab} \delta \bar{T}_{ab}^{\text{loc}} = {{NQ} \over {16}} \Big(\alpha
 + {1 \over \alpha}\Big) \left( 3\pi_1 + \pi_2 \right) \,,
} where we have used $2 \kappa_{10}^2
 T_3 = (2 \pi)^4 \alpha'^2$.

Now we equate $\delta R_{MN}$ with $\delta\bar{T}_{MN}$ and expand
each fluctuation mode in terms of the spherical harmonics. The
trace part of (\ref{Ricci1}) and (\ref{emtensor1}) then reads
\begin{equation}\begin{split}
\bigg[ & \Big\{  \square_x-\tfrac23\square_y + \square_z -
\tfrac{13}8
\Big(\alpha+\frac{1}{\alpha}\Big)NQ\Big\}\pi_{(1)}^{IJ}+\Big\{-\square_y+\tfrac98\Big(\alpha+\frac{1}{\alpha}\Big)
NQ\Big\}\pi_{(2)}^{IJ} \\& +  \square_y
H^{IJ}-3NQ\Big(\alpha-\frac{1}{\alpha}\Big)\delta\phi^{IJ} + 6
\alpha N^2 \delta C_0^{IJ} +\frac{Q}{\alpha N}\big(\tfrac92 N_1
\square_y b_{(1)}^{IJ}-\tfrac32 N_2 \square_z b_{(2)}^{IJ}\big)
\\ & + \frac{\alpha N}{Q}\big(\tfrac92 Q_1 \square_y
c_{(1)}^{IJ}-\tfrac32 Q_2 \square_z
c_{(2)}^{IJ}\big)\bigg]Y^IZ^J=0 \,, \label{finalmetric11}
\end{split}\end{equation}
while the traceless part $\delta R_{(ab)}=\delta\bar{T}_{(ab)}$
gives the following equation:
\begin{gather}
\big(H^{IJ}-\tfrac53\pi_{(1)}^{IJ}-\pi_{(2)}^{IJ}\big)\nabla_{(a}\nabla_{b)}Y^IZ^J=0
\,. \label{finalmetric12}
\end{gather}
We get similar set of equations from (\ref{Ricci2}) and
(\ref{emtensor2}):
\begin{equation}\begin{split}
\bigg[ & \Big\{  \square_x+\square_y - \tfrac23\square_z
-\tfrac{13}8
\Big(\alpha+\frac{1}{\alpha}\Big)NQ\Big\}\pi_{(2)}^{IJ}+\Big\{-\square_z+\tfrac98\Big(\alpha+\frac{1}{\alpha}\Big)
NQ\Big\}\pi_{(1)}^{IJ} \\ & + \square_z
H^{IJ}+3NQ\Big(\alpha-\frac{1}{\alpha}\Big)\delta\phi^{IJ} - 6
\alpha N^2 \delta C_0^{IJ} -\frac{Q}{\alpha N}\big(\tfrac32 N_1
\square_y b_{(1)}^{IJ}-\tfrac92 N_2 \square_z b_{(2)}^{IJ}\big) \\
&- \frac{\alpha N}{Q}\big(\tfrac32 Q_1 \square_y
c_{(1)}^{IJ}-\tfrac92 Q_2 \square_z
c_{(2)}^{IJ}\big)\bigg]Y^IZ^J=0\,, \label{finalmetric21}
\end{split}\end{equation}
and
\begin{gather}
\big(H^{IJ}-\pi_{(1)}^{IJ}-\tfrac53\pi_{(2)}^{IJ}\big)Y^I\nabla_{(i}\nabla_{j)}Z^J=0\,.
\label{finalmetric22}
\end{gather}

\subsection{Mass Spectrum}
So far in this section, we derived the fluctuation equations of
motion assuming that the internal manifold is $A_3\times B_3$
without the orientifold action. But as we saw in Section
\ref{Realistic}, the tadpole constraint (\ref{tadpole}) requires
the presence of O3-planes in the solution. We now discuss how this
will affect the mass spectrum.

The orientifold action can be written as $\Omega (-1)^{F_L}R$,
where $R$ is a $\mathbf{Z}_2$ action on the internal manifold and
$\Omega (-1)^{F_L}$ takes the value $+1$ on the metric $g_{MN}$,
R-R 4-form $C_4$, axion $C_0$ and dilaton $\phi$, and $-1$ on the
NS-NS and R-R 2-form potentials $B_2$ and $C_2$ \cite{FP,KST}.
This implies that we want our metric, axion, and dilaton
fluctuations to transform by $+1$ under the $\mathbf{Z}_2$ action
$R$, and NS-NS and R-R 2-form fluctuations to transform by $-1$.
We shall consider a specific type of $\mathbf{Z}_2$ action
$R=R_1\times R_2:(y,z)\mapsto(y',z')$ on $A_3\times B_3$, where
$R_1:y\mapsto y'$ and $R_2:z\mapsto z'$ are $\mathbf{Z}_2$
symmetries of $A_3$ and and $B_3$, respectively. Then our
background metric (\ref{ProdSpace}) and dilaton expectation value
(\ref{dilback}) are obviously consistent with the orientifold
action.

To see the consistency of our background $H_3$ and $F_3$ choices,
we need to consider how they transform under the $\mathbf{Z}_2$
action $R$. Let us first consider
$H_{abc}(y)=N_1\epsilon_{abc}(y)$. The transformation property of
$B_{ab}$ under $R$ can be written as
 \begin{equation}\label{B2transform}
 B_{ab}(y')=-\frac{\partial y^{c}}{\partial y^{'a}}\frac{\partial
 y^{d}}{\partial y^{'b}}B_{cd}(y)\,.
 \end{equation}
This implies
 \begin{equation}\label{H3transform}\begin{split}
 H_{abc}(y')&=\partial_a B_{bc}(y')+\text{cyclic permutations in
 $a,b,c$} \\
 &=\frac{\partial}{\partial y^{'a}}\left[-\frac{\partial y^{d}}{\partial y^{'b}}\frac{\partial
 y^{e}}{\partial y^{'c}}B_{de}(y)\right]+\text{permutations} \\
 &=-\frac{\partial y^d}{\partial y^{'b}}\frac{\partial
 y^e}{\partial y^{'c}}\frac{\partial y^f}{\partial
 y^{'a}}\frac{\partial B_{de}(y)}{\partial
 y^f}+\text{permutations} \\
 &=-\frac{\partial y^f}{\partial y^{'a}}\frac{\partial
 y^d}{\partial y^{'b}}\frac{\partial y^e}{\partial
 y^{'c}}H_{fde}(y)=-\big(\pm N_1\epsilon_{abc}(y')\big)\,,
 \end{split}\end{equation}
where the sign in the last line depends on whether $R_1$ is
orientation-preserving or not. Therefore, we see that the
background $H_{abc}(y)=N_1\epsilon_{abc}(y)$ is consistent with
the orientifold action only if $R_1$ is an orientation-reversing
map on $A_3$. Similarly, $H_{ijk}=N_2\epsilon_{ijk}(z)$ is a
consistent background only if $R_2$ is orientation-reversing on
$B_3$. The same holds true for $F_3$ background values.

Having shown that both $R_1$ and $R_2$ must be
orientation-reversing maps on respective manifolds, we determine which
fluctuation modes will survive. We can choose
our spherical harmonics $Y^I(y)$ and $Z^J(z)$ to be the parity
eigenmodes under $R_1$ and $R_2$:
 \begin{equation}
 Y^I(y')=\pm Y^I(y)\,,\quad Z^J(z')=\pm Z^J(z)\,.
 \end{equation}
We shall denote the parity eigenvalues by $(-1)^I$ and $(-1)^J$,
respectively, and call the indices $I$ and $J$ even or odd
depending on the parity of $Y^I(y)$ and $Z^J(z)$. From the mode
expansion given in Eqs.~(\ref{Taufluct})--(\ref{adsfluct}), it is
then clear that for the fluctuations $H^{IJ}$,
$\pi_{(1),(2)}^{IJ}$, $\delta\phi^{IJ}$, and $\delta C_0^{IJ}$,
the modes that survive the orientifold projection are those with
$I$ and $J$ both even or both odd. Things are slightly trickier
for the 2-form potential fluctuations $b_2$ and $c_2$. For
example, let's consider the fluctuation $b_{ab}$, whose mode
expansion is given in Eq.~(\ref{b2fluct}). Under the action of
$R$, it transforms as
 \begin{equation}\begin{split}
 b_{(1)}^{IJ}(x)\epsilon_{ab}^{\;\;\;\;c}(y)&\nabla_cY^I(y)Z^J(z)  \rightarrow
  \frac{\partial y^{'c}}{\partial y^a}\frac{\partial y^{'d}}{\partial y^b}
  b_{(1)}^{IJ}(x){\epsilon}_{cd}^{\;\;\;\;e}(y')\nabla_eY^I(y')Z^J(z') \\
  & =b_{(1)}^{IJ}(x)\frac{\partial y^{'c}}{\partial y^a}\frac{\partial
  y^{'d}}{\partial
  y^b}{\epsilon}_{cd}^{\;\;\;\;e}(y')\frac{\partial y^f}{\partial
  y^{'e}}\frac{\partial}{\partial y^f}(-1)^IY(y)(-1)^JZ^J(z)\\
  & =-b_{(1)}^{IJ}(x)\epsilon_{ab}^{\;\;\;\;f}(y)(-1)^I\nabla_fY(y)(-1)^JZ^J(z)\,,
 \end{split}\end{equation}
where the minus sign in the third line comes from the fact that
$R_1$ is orientation-reversing. Since the R-R 2-form should
transform by $-1$ under $R$, the modes $b_{(1)}^{IJ}$ that survive
the projection are those with $I$ and $J$ both even or both odd.
The cases for $b_{ij}$, $c_{ab}$, and $c_{ij}$ are all exactly the
same. In conclusion, for all the fluctuation modes given in
Eqs.~(\ref{Taufluct})--(\ref{adsfluct}), those which survive the
orientifold projection are precisely those with $I$ and $J$ both
even or both odd.

Now we compute the spectrum of the fluctuation modes that survive
the orientifold projection and determine the stability of our
background. Let us first introduce some notations. We denote the
eigenvalues of the Laplacians on $A_3$ and $B_3$ as follows:
 \begin{equation}
 \square_y Y^I=-\frac{\lambda^I}{R^2}Y^I\,,\quad \square_z
 Z^J=-\frac{\mu^J}{R^2}Z^J\,.
 \end{equation}
For the lowest harmonics on the respective manifolds, that is,
$Y^I(y)=\text{constant}$ and $Z^J(z)=\text{constant}$, the
corresponding eigenvalues are obviously $\lambda^I=\mu^J=0$. For
non-constant $Y^I$ and $Z^J$, they have a lower bound
$\lambda^I,\mu^J \geq 3$, which holds for any 3-dimensional
compact Einstein manifold of positive curvature \cite{DNP}. We also find it convenient to re-scale the 2-form
potential fluctuations as follows:
 \begin{equation}\begin{split}
 b_{(1)}^{IJ}&=\tfrac{1}{\alpha}N_1L^2B_{(1)}^{IJ}\,,\qquad b_{(2)}=\alpha
 N_2L^2B_{(2)}^{IJ}\,, \\  c_{(1)}^{IJ}&=\alpha
 Q_1L^2C_{(1)}^{IJ}\,,\qquad
 c_{(2)}^{IJ}=\tfrac{1}{\alpha}Q_2L^2C_{(2)}^{IJ}\,.
 \end{split}\end{equation}

We start with the lowest harmonics on both manifolds. From
Eqs.~(\ref{b2fluct}) and (\ref{c2fluct}), we see that there are no
form field fluctuations at this level, as they involve derivatives
on $Y^I$ and $Z^J$. But all the other fluctuations survive
orientifold projection because $Y^I$ and $Z^J$, both being
constant, are obviously even under the $\mathbf{Z}_2$ action.
Using the fact that $\square_yY^I=0$ and $\square_zZ^J=0$, the
contents of dilaton equation (\ref{finaldilaton}), axion equation
(\ref{finalaxion}), and metric equations (\ref{finalmetric11}) and
(\ref{finalmetric21}) can be summarized as follows:
 \begin{equation}
 L^2\square_x \left(\begin{array}{c} \delta\phi \\ \delta C_0 \\
 \pi_{(1)} \\ \pi_{(2)} \end{array}\right) = \left(
 \begin{array}{cccc} 24 & 0 & \tfrac{12(\alpha^2-1)}{\alpha^2+1} & -\tfrac{12(\alpha^2-1)}{\alpha^2+1} \\
 0 & 24 & -\tfrac{24\alpha}{(\alpha^2+1)g_s} & \tfrac{24\alpha}{(\alpha^2+1)g_s} \\ \tfrac{72(\alpha^2-1)}{\alpha^2+1} &
 -\tfrac{144\alpha g_s}{\alpha^2+1} & 39 & -27 \\
 -\tfrac{72(\alpha^2-1)}{\alpha^2+1} &
 \tfrac{144\alpha g_s}{\alpha^2+1} & -27 & 39 \end{array}\right)\left(\begin{array}{c} \delta\phi \\ \delta C_0 \\
 \pi_{(1)} \\ \pi_{(2)} \end{array}\right) \,.
 \end{equation}
Here we have used the relations (\ref{dilback}) and
(\ref{adsradius}) for our background. All the other fluctuation
equations of motion are trivially satisfied because
$\nabla_aY^I=\nabla_iZ^J=0$. The eigenvalues of the mass matrix
are
$$m^2L^2=12,\,24,\,3(15 \pm \sqrt{241})\,.$$
Even though one of them has negative mass-squared, they are all
above the BF bound (\ref{Bound}), and the background is stable
against this fluctuation mode.

Next, we consider the modes where one of the harmonics $Y^I$ and
$Z^J$ is constant but the other is a higher harmonic. For
definiteness, let us assume that $\nabla_aY^I=0$ but
$\nabla_iZ^J\neq 0$. Then the fluctuation $b_{(1)}^{IJ}$ and
$c_{(1)}^{IJ}$ are still absent, but we now have $b_{(2)}^{IJ}$
and $c_{(2)}^{IJ}$ for even $J$ (because $I$ is even). All the
other fluctuations are also present for even $J$. Since
$\nabla_{(i}\nabla_{j)}Z^J\neq 0$ for nonconstant $Z^J$ on a
generic\footnote{This does not hold when $B_3$ is maximally
symmetric, i.e. a 3-sphere. In this case, the eigenvalues of the
Laplacian are given by $\mu^J=k(k+2)$, where $k=0,1,2,\ldots$, and
$\nabla_{(i}\nabla_{j)}Z^J=0$ not only for $k=0$ mode but also for
$k=1$ modes. Therefore, the algebraic constraint in
(\ref{constraint}) holds only for $k\geq 2$ modes.} 3-manifold
$B_3$, Eq.~(\ref{finalmetric22}) gives an algebraic constraint
 \begin{equation}\label{constraint}
 H^{IJ}-\pi_{(1)}^{IJ}-\tfrac53\pi_{(2)}^{IJ}=0\,,
 \end{equation}
which we can use to eliminate $H^{IJ}$ in
Eq.~(\ref{finalmetric21}) in favor of $\pi_{(1)}$ and $\pi_{(2)}$.
Using $\square_y Y^I=0$ and $\square_z Z^J=-(\mu^J/R^2)Z^J$, the
fluctuation equations (\ref{finaldilaton}), (\ref{finalaxion}),
(\ref{finalh32}), (\ref{finalf32}), (\ref{finalmetric11}), and
(\ref{finalmetric21}) result in the following mass matrix:
 \begin{equation}\label{mmb2}
 L^2\square_x \left(\begin{array}{c} \delta\phi \\ \delta C_0 \\
 B_{(2)} \\ C_{(2)} \\ \pi_{(1)} \\ \pi_{(2)} \end{array}\right) = M^2
 \left(\begin{array}{c} \delta\phi \\ \delta C_0 \\
 B_{(2)} \\ C_{(2)} \\ \pi_{(1)} \\ \pi_{(2)} \end{array}\right) \,,
 \end{equation}
where
 \begin{equation} M^2=\left( \begin{array}{cccccc} \tfrac32\mu^J+24 & 0 &
 \tfrac{36\alpha\mu^J}{\alpha^2+1} &
 -\tfrac{36\alpha\mu^J}{\alpha^2+1} & \tfrac{12
 (\alpha^2-1)}{\alpha^2+1} & -\tfrac{12
 (\alpha^2-1)}{\alpha^2+1} \\ 0 & \tfrac32 \mu^J +24 & \tfrac{36\alpha^2\mu^J}{(\alpha^2+1)g_s} &
 \tfrac{36\mu^J}{(\alpha^2+1)g_s} & -\tfrac{24\alpha}{(\alpha^2+1)g_s} &
 \tfrac{24\alpha}{(\alpha^2+1)g_s} \\ \tfrac{1}{\alpha} & g_s
 & \tfrac32\mu^J & 0& 0 & \tfrac{2}{3\alpha} \\ -\alpha &
 g_s & 0 & \tfrac32 \mu^J & 0 & \tfrac{2\alpha}{3} \\
 \tfrac{72 (\alpha^2-1)}{\alpha^2+1} &
 -\tfrac{144\alpha g_s}{\alpha^2+1} & -\tfrac{54\alpha
 \mu^J}{\alpha^2+1} & -\tfrac{54\alpha\mu^J}{\alpha^2+1} &
 \tfrac32\mu^J+39 & -27 \\ -\tfrac{72 (\alpha^2-1)}{\alpha^2+1} &
 \tfrac{144\alpha g_s}{\alpha^2+1} & \tfrac{162\alpha
 \mu^J}{\alpha^2+1} & \tfrac{162\alpha\mu^J}{\alpha^2+1}
 & -27 & \tfrac32\mu^J+39 \end{array}\right)\,. \label{mmb22}
 \end{equation}
The expression for the eigenvalues of (\ref{mmb22}) in terms of
$\mu^J$ is quite complicated, but it is independent of $\alpha$ and $g_s$.
One also finds that the smallest eigenvalue lies below the BF
bound if $3\leq \mu^J \lesssim 22.1$. That is, the background is
unstable if $B_3$ has a non-constant {\it even} scalar harmonic with
$\mu^J\lesssim 22.1$. Exchanging the roles of $A_3$ and $B_3$
yields exactly same result: the background is unstable if $A_3$
has a non-constant scalar harmonic with $\lambda^I\lesssim 22.1$.

Assuming that non-zero eigenvalues of the Laplacian on $A_3$ and
$B_3$ are all larger than $\sim 22.1$, we continue to check the
modes with both $Y^I$ and $Z^J$ non-constant. For $I$ and $J$ both
even or both odd, we obtain another algebraic constraint from
Eq.~(\ref{finalmetric22}) which is analogous to
Eq.~(\ref{constraint}). Together, these two constraints imply
 \begin{equation}\label{constraint2}
 \pi_{(1)}^{IJ}=\pi_{(2)}^{IJ}\equiv \pi^{IJ}\quad\text{and}\quad
 H^{IJ}=\tfrac83\pi^{IJ}\,,
 \end{equation}
which allows us to eliminate $H^{IJ}$ and $\pi_{(1),(2)}^{IJ}$ in
favor of $\pi^{IJ}$. The mass matrix we get from the fluctuation
equations is then
\begin{equation}
 L^2\square_x\left(\begin{array}{c} \delta\phi \\ \delta C_0 \\
 B_{(1)} \\ B_{(2)} \\ C_{(1)} \\ C_{(2)} \\ \pi
 \end{array}\right)= M^2 \left(\begin{array}{c} \delta\phi \\ \delta C_0 \\
 B_{(1)} \\ B_{(2)} \\ C_{(1)} \\ C_{(2)} \\ \pi
 \end{array}\right) \,,
 \end{equation}
where $M^2$ is now
 \begin{equation*}
 \left(\begin{array}{ccccccc}
 \tfrac{3(\lambda^I+\mu^J)+48}{2} & 0 &
 \tfrac{36\alpha\lambda^I}{\alpha^2+1} & \tfrac{36\alpha\mu^J}{\alpha^2+1}
 & -\tfrac{36\alpha\lambda^I}{\alpha^2+1} & -\tfrac{36\alpha\mu^J}{\alpha^2+1}
 & 0 \\ 0 & \tfrac{3(\lambda^I+\mu^J)+48}{2} & -\tfrac{36\lambda^I}{(\alpha^2+1)g_s}
 & \tfrac{36\alpha^2\mu^J}{(\alpha^2+1)g_s}
 & -\tfrac{36\alpha^2\lambda^I}{(\alpha^2+1)g_s} & \tfrac{36\mu^J}{(\alpha^2+1)g_s}
 & 0 \\ \alpha & -g_s & \tfrac{3(\lambda^I+\mu^J)}{2} & 0
 & 0 & 0 & \tfrac{2\alpha}{3} \\ \tfrac{1}{\alpha} & g_s &
 0 & \tfrac{3(\lambda^I+\mu^J)}{2} & 0 & 0 & \tfrac{2}{3\alpha} \\
 -\tfrac{1}{\alpha} & -g_s & 0 & 0 & \tfrac{3(\lambda^I+\mu^J)}{2} & 0
 & \tfrac{2}{3\alpha} \\ -\alpha & g_s & 0 & 0 & 0 &
 \tfrac{3(\lambda^I+\mu^J)}{2}
 & \tfrac{2\alpha}{3} \\ 0 & 0 &
 \tfrac{54\alpha\lambda^I}{\alpha^2+1} &
 \tfrac{54\alpha\mu^J}{\alpha^2+1}&
 \tfrac{54\alpha\lambda^I}{\alpha^2+1}&
 \tfrac{54\alpha\mu^J}{\alpha^2+1}&
 \tfrac{3(\lambda^I+\mu^J)+24}{2} \end{array}\right)\,.
 \end{equation*}
Again, the eigenvalues are quite complicated functions of
$\lambda^I$ and $\mu^J$, but numerical experimentation shows that
they are all above the BF bound for $\lambda^I,\mu^J\gtrsim 22.1$.

From these observations, we conclude that the criterion for the
stability of these Freund--Rubin compactifications is that there
be no {\it even} eigenfunction of the scalar Laplacian with
eigenvalue between $3$ and $22.1$.

We end this section with the simplest example of this kind where
we take both $A_3$ and $B_3$ to be 3-spheres. For each 3-sphere,
we take the coordinate system $(\psi,\theta,\phi)$ defined by
 \begin{alignat*}{2}
 x&=R\cos\psi\,,&\qquad& y=R\sin\psi\cos\theta\,,\\
 z&=R\sin\psi\sin\theta\cos\phi\,,
 && w=R\sin\psi\sin\theta\sin\phi\,,
 \end{alignat*}
in terms of the 3-sphere $x^2+y^2+z^2+w^2=R^2$ embedded in
$\mathbb{R}^4$. The orientation-reversing $\mathbf{Z}_2$ action on
$S^3$ is defined as
 \begin{equation}\label{Zexample}
 \psi\rightarrow\pi-\psi\,,\quad
 \theta\rightarrow\pi-\theta\,,\quad \phi\rightarrow 2\pi-\phi\,.
 \end{equation}
It has two fixed points on each 3-sphere, so the number of
O3-planes on $S^3\times S^3$ is four. The eigenvalues of the
Laplacian on $S^3$ are given by $\mu^J=k(k+2)$ where
$k=0,1,2,\ldots$, and we obtain the mass matrix Eq.~(\ref{mmb2})
when $k=0$ on the first 3-sphere and $k\geq 2$ on the second. In
particular, for $k=2$ and $k=3$ on the second 3-sphere, the
corresponding eigenvalues are $\mu^J=8$ and 15. Since both $k=2$
and $k=3$ have modes which are even under (\ref{Zexample}), this
background is unstable against small fluctuations.

\section*{Acknowledgements}

We would like to thank M.~Douglas, O.~Ganor, P.~Horava, S.~Kachru, I.~Klebanov,
 L.~McAllister, J.~McGreevy, W.~Taylor and especially
 O.~deWolfe and E.~Silverstein for helpful discussions. We
 would also like to thank T.~Shiromizu for a helpful
 correspondence. Y.~P.~H would like to thank the organizers and
 participants of TASI 2005 for hospitality and helpful discussions.
I.~M. was
supported in part by the Director, Office of Science, Office of High
Energy and Nuclear Physics, of the U.S. Department of Energy under
Contract ~DE-AC02-05CH11231, in part by the National Science
 Foundation under grant PHY-00-98840, and by the Berkeley Center for
 Theoretical Physics. I.~M. would also like to thank
 the Aspen Center for Physics for hospitality during the final stages
 of this work.

\bibliographystyle{ssg}
\bibliography{landscape14}

\begingroup\raggedright\begin{thebibliography}{10}

\bibitem{KKLT}
S.~Kachru, R.~Kallosh, A.~Linde, and S.~P. Trivedi, ``De Sitter vacua in string
  theory,'' {\em Phys. Rev.} {\bf D68} (2003) 046005,
  \href{http://xxx.lanl.gov/abs/hep-th/0301240}{{\tt hep-th/0301240}}.

\bibitem{DGKT}
O.~DeWolfe, A.~Giryavets, S.~Kachru, and W.~Taylor, ``Type IIA moduli
  stabilization,'' {\em JHEP} {\bf 07} (2005) 066,
  \href{http://xxx.lanl.gov/abs/hep-th/0505160}{{\tt hep-th/0505160}}.

\bibitem{DGKTTwo}
O.~DeWolfe, A.~Giryavets, S.~Kachru, and W.~Taylor, ``Enumerating flux vacua
  with enhanced symmetries,'' {\em JHEP} {\bf 02} (2005) 037,
  \href{http://xxx.lanl.gov/abs/hep-th/0411061}{{\tt hep-th/0411061}}.

\bibitem{Oliver}
O.~DeWolfe, ``Enhanced symmetries in multiparameter flux vacua,''
  \href{http://xxx.lanl.gov/abs/hep-th/0506245}{{\tt hep-th/0506245}}.

\bibitem{FFix}
F.~Denef, M.~R. Douglas, B.~Florea, A.~Grassi, and S.~Kachru, ``Fixing all
  moduli in a simple F-theory compactification,''
  \href{http://xxx.lanl.gov/abs/hep-th/0503124}{{\tt hep-th/0503124}}.

\bibitem{Conlon}
J.~P. Conlon, F.~Quevedo, and K.~Suruliz, ``Large-volume flux
  compactifications: Moduli spectrum and D3/D7 soft supersymmetry breaking,''
  \href{http://xxx.lanl.gov/abs/hep-th/0505076}{{\tt hep-th/0505076}}.

\bibitem{Vijay}
V.~Balasubramanian, P.~Berglund, J.~P. Conlon, and F.~Quevedo, ``Systematics of
  moduli stabilisation in Calabi-Yau flux compactifications,'' {\em JHEP} {\bf
  03} (2005) 007, \href{http://xxx.lanl.gov/abs/hep-th/0502058}{{\tt
  hep-th/0502058}}.

\bibitem{Acharya}
B.~S. Acharya, ``A moduli fixing mechanism in M theory,''
  \href{http://xxx.lanl.gov/abs/hep-th/0212294}{{\tt hep-th/0212294}}.

\bibitem{Curio}
G.~Curio, A.~Krause, and D.~Lust, ``Moduli stabilization in the heterotic / IIB
  discretuum,'' \href{http://xxx.lanl.gov/abs/hep-th/0502168}{{\tt
  hep-th/0502168}}.

\bibitem{BeckOne}
K.~Becker, M.~Becker, K.~Dasgupta, and P.~S. Green, ``Compactifications of
  heterotic theory on non-Kaehler complex manifolds. I,'' {\em JHEP} {\bf 04}
  (2003) 007, \href{http://xxx.lanl.gov/abs/hep-th/0301161}{{\tt
  hep-th/0301161}}.

\bibitem{BeckTwo}
K.~Becker, M.~Becker, P.~S. Green, K.~Dasgupta, and E.~Sharpe,
  ``Compactifications of heterotic strings on non-Kaehler complex manifolds.
  II,'' {\em Nucl. Phys.} {\bf B678} (2004) 19--100,
  \href{http://xxx.lanl.gov/abs/hep-th/0310058}{{\tt hep-th/0310058}}.

\bibitem{BA}
R.~Brustein and S.~P. de~Alwis, ``Moduli potentials in string compactifications
  with fluxes: Mapping the discretuum,'' {\em Phys. Rev.} {\bf D69} (2004)
  126006, \href{http://xxx.lanl.gov/abs/hep-th/0402088}{{\tt hep-th/0402088}}.

\bibitem{STW}
J.~Shelton, W.~Taylor, and B.~Wecht, ``Nongeometric Flux Compactifications,''
  \href{http://xxx.lanl.gov/abs/hep-th/0508133}{{\tt hep-th/0508133}}.

\bibitem{AMOne}
I.~Antoniadis and T.~Maillard, ``Moduli stabilization from magnetic fluxes in
  type I string theory,'' {\em Nucl. Phys.} {\bf B716} (2005) 3--32,
  \href{http://xxx.lanl.gov/abs/hep-th/0412008}{{\tt hep-th/0412008}}.

\bibitem{AMTwo}
I.~Antoniadis, A.~Kumar, and T.~Maillard, ``Moduli stabilization with open and
  closed string fluxes,'' \href{http://xxx.lanl.gov/abs/hep-th/0505260}{{\tt
  hep-th/0505260}}.

\bibitem{Krause}
G.~Curio and A.~Krause, ``G-fluxes and non-perturbative stabilisation of
  heterotic M- theory,'' {\em Nucl. Phys.} {\bf B643} (2002) 131--156,
  \href{http://xxx.lanl.gov/abs/hep-th/0108220}{{\tt hep-th/0108220}}.

\bibitem{BoussoPol}
R.~Bousso and J.~Polchinski, ``Quantization of four-form fluxes and dynamical
  neutralization of the cosmological constant,'' {\em JHEP} {\bf 06} (2000)
  006, \href{http://xxx.lanl.gov/abs/hep-th/0004134}{{\tt hep-th/0004134}}.

\bibitem{Douglas}
M.~R. Douglas, ``The statistics of string / M theory vacua,'' {\em JHEP} {\bf
  05} (2003) 046, \href{http://xxx.lanl.gov/abs/hep-th/0303194}{{\tt
  hep-th/0303194}}.

\bibitem{DA}
S.~Ashok and M.~R. Douglas, ``Counting flux vacua,'' {\em JHEP} {\bf 01} (2004)
  060, \href{http://xxx.lanl.gov/abs/hep-th/0307049}{{\tt hep-th/0307049}}.

\bibitem{DD}
F.~Denef and M.~R. Douglas, ``Distributions of flux vacua,'' {\em JHEP} {\bf
  05} (2004) 072, \href{http://xxx.lanl.gov/abs/hep-th/0404116}{{\tt
  hep-th/0404116}}.

\bibitem{Land}
L.~Susskind, ``The anthropic landscape of string theory,''
  \href{http://xxx.lanl.gov/abs/hep-th/0302219}{{\tt hep-th/0302219}}.

\bibitem{DNONSUSY}
M.~R. Douglas, ``Statistical analysis of the supersymmetry breaking scale,''
  \href{http://xxx.lanl.gov/abs/hep-th/0405279}{{\tt hep-th/0405279}}.

\bibitem{BDM}
R.~Bousso, O.~DeWolfe, and R.~C. Myers, ``Unbounded entropy in spacetimes with
  positive cosmological constant,'' {\em Found. Phys.} {\bf 33} (2003)
  297--321, \href{http://xxx.lanl.gov/abs/hep-th/0205080}{{\tt
  hep-th/0205080}}.

\bibitem{KPZ}
C.~Krishnan, S.~Paban, and M.~Zanic, ``Evolution of gravitationally unstable de
  Sitter compactifications,'' {\em JHEP} {\bf 05} (2005) 045,
  \href{http://xxx.lanl.gov/abs/hep-th/0503025}{{\tt hep-th/0503025}}.

\bibitem{FR}
P.~G.~O. Freund and M.~A. Rubin, ``Dynamics of Dimensional Reduction,'' {\em
  Phys. Lett.} {\bf B97} (1980) 233--235.

\bibitem{EvaModuli}
E.~Silverstein, ``TASI / PiTP / ISS lectures on moduli and microphysics,''
  \href{http://xxx.lanl.gov/abs/hep-th/0405068}{{\tt hep-th/0405068}}.

\bibitem{KKLLMMT}
S.~Kachru {\em et.~al.}, ``Towards inflation in string theory,'' {\em JCAP}
  {\bf 0310} (2003) 013, \href{http://xxx.lanl.gov/abs/hep-th/0308055}{{\tt
  hep-th/0308055}}.

\bibitem{Campbell}
I.~C.~G. Campbell, ``The Stability of ten-dimensional Kaluza-Klein
  Supergravity,'' {\em Phys. Rev.} {\bf D31} (1985) 1911.

\bibitem{SW}
L.~Susskind and E.~Witten, ``The holographic bound in anti-de Sitter space,''
  \href{http://xxx.lanl.gov/abs/hep-th/9805114}{{\tt hep-th/9805114}}.

\bibitem{GiantInf}
O.~DeWolfe, S.~Kachru, and H.~L. Verlinde, ``The giant inflaton,'' {\em JHEP}
  {\bf 05} (2004) 017, \href{http://xxx.lanl.gov/abs/hep-th/0403123}{{\tt
  hep-th/0403123}}.

\bibitem{Sen}
A.~Sen, ``F-theory and Orientifolds,'' {\em Nucl. Phys.} {\bf B475} (1996)
  562--578, \href{http://xxx.lanl.gov/abs/hep-th/9605150}{{\tt
  hep-th/9605150}}.

\bibitem{FP}
A.~R. Frey and J.~Polchinski, ``N = 3 warped compactifications,'' {\em Phys.
  Rev.} {\bf D65} (2002) 126009,
  \href{http://xxx.lanl.gov/abs/hep-th/0201029}{{\tt hep-th/0201029}}.

\bibitem{KST}
S.~Kachru, M.~B. Schulz, and S.~Trivedi, ``Moduli stabilization from fluxes in
  a simple IIB orientifold,'' {\em JHEP} {\bf 10} (2003) 007,
  \href{http://xxx.lanl.gov/abs/hep-th/0201028}{{\tt hep-th/0201028}}.

\bibitem{BF}
P.~Breitenlohner and D.~Z. Freedman, ``Stability in Gauged Extended
  Supergravity,'' {\em Ann. Phys.} {\bf 144} (1982) 249.

\bibitem{Duff}
M.~J. Duff, B.~E.~W. Nilsson, and C.~N. Pope, ``THE Criterion for Vacuum
  Stability in Kaluza-Klein Supergravity,'' {\em Phys. Lett.} {\bf B139} (1984)
  154.

\bibitem{Cosets}
P.~van Nieuwenhuizen, ``The Complete Mass Spectrum of d = 11 Supergravity
  Compactified on S(4) and a General Mass Formula for Arbitrary Cosets M(4),''
  {\em Class. Quant. Grav.} {\bf 2} (1985) 1.

\bibitem{Biran}
B.~Biran, A.~Casher, F.~Englert, M.~Rooman, and P.~Spindel, ``The Fluctuating
  Seven Sphere in Eleven-Dimensional Supergravity,'' {\em Phys. Lett.} {\bf
  B134} (1984) 179.

\bibitem{DFGHM}
O.~DeWolfe, D.~Z. Freedman, S.~S. Gubser, G.~T. Horowitz, and I.~Mitra,
  ``Stability of AdS(p) x M(q) compactifications without supersymmetry,'' {\em
  Phys. Rev.} {\bf D65} (2002) 064033,
  \href{http://xxx.lanl.gov/abs/hep-th/0105047}{{\tt hep-th/0105047}}.

\bibitem{Heterotic}
S.~Gukov, S.~Kachru, X.~Liu, and L.~McAllister, ``Heterotic moduli
  stabilization with fractional Chern-Simons invariants,'' {\em Phys. Rev.}
  {\bf D69} (2004) 086008, \href{http://xxx.lanl.gov/abs/hep-th/0310159}{{\tt
  hep-th/0310159}}.

\bibitem{Shiromizu}
T.~Shiromizu, D.~Ida, H.~Ochiai, and T.~Torii, ``Stability of AdS(p) x S(n) x
  S(q-n) compactifications,'' {\em Phys. Rev.} {\bf D64} (2001) 084025,
  \href{http://xxx.lanl.gov/abs/hep-th/0106265}{{\tt hep-th/0106265}}.

\bibitem{Polbook}
J.~Polchinski, ``String theory. Vol. 2: Superstring theory and beyond,''.
  Cambridge, UK: Univ. Pr. (1998) 531 p.

\bibitem{GKP}
S.~B. Giddings, S.~Kachru, and J.~Polchinski, ``Hierarchies from fluxes in
  string compactifications,'' {\em Phys. Rev.} {\bf D66} (2002) 106006,
  \href{http://xxx.lanl.gov/abs/hep-th/0105097}{{\tt hep-th/0105097}}.

\bibitem{GP}
M.~Grana and J.~Polchinski, ``Supersymmetric three-form flux perturbations on
  AdS(5),'' {\em Phys. Rev.} {\bf D63} (2001) 026001,
  \href{http://xxx.lanl.gov/abs/hep-th/0009211}{{\tt hep-th/0009211}}.

\bibitem{SSG}
S.~S. Gubser, ``Supersymmetry and F-theory realization of the deformed conifold
  with three-form flux,'' \href{http://xxx.lanl.gov/abs/hep-th/0010010}{{\tt
  hep-th/0010010}}.

\bibitem{Junction}
E.~Silverstein, ``AdS and dS entropy from string junctions or the function of
  junction conjunctions,'' \href{http://xxx.lanl.gov/abs/hep-th/0308175}{{\tt
  hep-th/0308175}}.

\bibitem{DNP}
M.~J. Duff, B.~E.~W. Nilsson, and C.~N. Pope, ``Kaluza-Klein Supergravity,''
  {\em Phys. Rept.} {\bf 130} (1986) 1--142.

\end{thebibliography}\endgroup

\end{document}